\documentclass[aps,prb,overcite,,twocolumn]{revtex4}
\usepackage{graphicx}       
\begin{document}      
\title{Negative refraction and left-handed behavior in two-dimensional photonic crystals}
\author{S. Foteinopoulou and C. M. Soukoulis}

\address{Ames Laboratory-USDOE and Department of Physics and
Astronomy,\\ Iowa State University, Ames, IA 50011\\}
\begin{abstract}{We systematically examine the conditions of obtaining left-handed (LH) behavior in photonic crystals. Detailed studies of the phase and group velocities as well as the phase $n_{p}$ and group $n_{g}$ refractive index are given. The existence of negative refraction does not guarantee the existence of negative index of refraction and so LH behavior. A wedge type of experiment is suggested that can unambiguously distiguinsh between cases of negative refraction that occur when left-handed behavior is present, from cases that show negative refraction without LH behavior.\\
PACS~numbers:~78.20.Ci, 41.20.Jb,42.25.-p, 42.30.-d \\}
\end{abstract}

\maketitle 
Recently, there have been many studies about materials that have a negative index of refraction $n$ \cite{smithpbg,smithprl,shelbyapl,shelbysc,mehmet}. These materials, theoretically discussed by Veselago\cite{vesalago}, have  simultaneous negative permittivity $\epsilon$ and permeability $\mu$. Pendry \cite{pendry,pendry00} has studied structures that they can have both $\epsilon$ and $\mu$ negative and suggested that they can be used to fabricate a perfect lens. In these materials, $\vec k$, $\vec E$, $\vec H$ form a left-handed (LH)  set of vectors (i.e., $\vec S \cdot \vec k <$ 0, where $\vec S$ is the Poynting vector) and, therefore, are called left-handed materials (LHM). Futhermore Marko\v{s} and Soukoulis \cite{markos} used the transfer matrix technique to calculate the transmission and reflection properties of left-handed materials. These scattering data were used \cite{smithprb} to determine the effective permitivity and permeability of the LHM and the index of refraction was found \cite{smithprb} to be unambiguously negative in the frequency region where both of the retrieved values of $\epsilon$ and $\mu$ are negative. All experiments \cite{smithpbg,smithprl,shelbyapl,shelbysc,mehmet} that showed left-handed behavior were performed in the microwave or millimeter regime.
\par
In addition Notomi \cite{Notomi} studied light propagation in strongly modulated two dimensional (2D) photonic crystals (PCs). Such a photonic crystal behaves as a material having an effective refractive index $n_{\rm eff}$ controllable by the band structure. In these PC structures the permittivity is periodically modulated in space and is positive. The permeability $\mu$ is equal to one. Negative $n_{\rm eff}$  for a frequency range was found.  The existence of negative $n$ was demonstrated \cite{Notomi} by a finite difference time domain (FDTD) simulation. Negative refraction on the interface of a 3D PC structure has been experimentally observed by Kosaka et al. \cite{kosaka}. A negative refractive index associated to the negative refraction was reported. Also in Ref. 12, large beam steering was observed that was called the ``superprism phenomenon''. Similar unusual light propagation was observed \cite{Russel} in 1D and 2D diffraction gratings. In addition, in Luo et. al. it was shown that the phenomenon of negative refraction in the PC can be utilized to make a superlens\cite{Luo}.  Finally a theoretical work \cite{Gralak} predicted a negative refraction index in PCs.         
\par 
Both in the PC  and LHM literature there is a lot of confusion about what are the correct definitions of the phase and group refractive index and what is their relation to negative refraction. It is important to examine if left-handed behavior can be observed in photonic crystals at optical frequencies. In addition, it is instructive to see how the latter relates with the sign of the phase and group refractive index for the PC system. 
\par

In this paper we will attempt to clarify the differences between phase refractive index $n_{p}$ and group refractive index $n_{g}$ in PC structures. We show that left-handed behavior in photonic crystals is possible and a comparison of this LH behavior to regular LH materials is given. We will also show that a frequency range exists for which negative refraction can occur while it does not correspond to regular LH behavior. A wedge type of experiment is suggested that can unveil the sign of $\vec S \cdot \vec k $\cite{note} (i.e. the ``rightness'') in the PC and so the correct sign for the ``effective'' index (phase index $n_{p}$ as we will see in the following). This way we can unambiguously distinguish between the case of ``negative refraction'' that corresponds to  LH behavior in the PC structure ($n_{p}$ and $\vec S \cdot \vec k $\cite{note}  are negative), from cases that show ``negative refraction'' without  LH behavior ($n_{p}$ and  $\vec S \cdot \vec k$ are positive). FDTD\cite{tavlove} simulations are presented which support our claims that left-handed behavior can be seen in the the PC only in the cases where the predicted value for $n_{p}$ (for the infinite system) is negative. In all FDTD simulations we used perfectly matched layer (PML) \cite{Berenger} boundary conditions.
\par 
Our results are general but for completeness and for comparison with the results of Notomi \cite{Notomi} we studied  a 2D hexagonal lattice of dielectric rods with dielectric constant 12.96 (e.g., GaAs or Si at 1.55 microns) for the H (TE) polarization. The radius of the dielectric rods is $r=0.35 a$, where $a$ is the lattice constant. In Fig. 1a the band structure of the system is shown. In the figure as well as in the following the quantity $\tilde{f}$ is the frequency normalized as $\omega a/2 \pi c$, where $c$ is the velocity of light. To study the system the surfaces (contours in the 2D system) of equal frequency in $\vec k$-space,-- called equifrequency surfaces (EFS)--, are needed. The EFS basically consists of the allowed propagation modes within the crystal for a certain frequency. For both the band structure and the EFS plots, the plane wave expansion method \cite{chan,villen} with 1003 plane waves was used. The EFS can have different shapes for various frequencies. We notice that in some cases their shape can be almost circular (i.e., $\omega \sim A k$, where $A$ a constant). Basically, these cases are for frequencies close to the band edges and we will refer to their dispersion relation $\omega(k)$ as normal dispersion from now on. \par
Before we proceed in presenting the FDTD simulation results it is useful to discuss  the appropriate expressions for the phase ($\vec{v}_{p}$) and group ($\vec{v}_{g}$) velocities and their associated phase ($n_{p}$) and group ($n_{g}$) refractive indices for the infinite photonic crystal system. The analysis that follows is not used in the interpretation of the FDTD results, that we discuss later in this paper. However it does  serve as a theoretical prediction to be compared with the FDTD data. For any general case $\vec{v}_{p}=\frac{c}{|n_{p}|} \hat {k} $ \hspace {0.01cm}, with $\hat {k}=\vec k/k$ \hspace {0.02cm}. Also \hspace {0.2cm} $|\vec{v}_{g}|$=$|\nabla_{k} \omega|$=$\frac{c}{|n_{g}|}$. The value of $|n_{p}|$ for a certain angle of incidence $\theta$ will be  $c|\vec{k}_{f}(\theta)|/\omega$ (the refracted wave-vector $\vec{k}_{f}$ can be determined from the EFS as it will be discussed later). The sign of $n_{p}$ is determined from the behavior of EFS. If the equifrequency contours move outwards with increasing frequency then $\vec{v}_{g} \cdot \vec{k}>$0; if they move inwards $\vec{v}_{g} \cdot \vec{k}<$0. It can be proven analytically that for the infinite PC system the group velocity coincides with the energy velocity \cite{fotein,sakoda}. Therefore the sign of $\vec{v}_{g} \cdot\vec{k}$ is equivalent to the sign of $\vec{S} \cdot\vec{k}$\cite{note} -- where $\vec k$ is in the $1^{st}$ BZ--. So, to be in accordance with the LH literature\cite{vesalago} the sign of $n_{p}$ will be the sign of $\vec{v}_{g} \cdot \vec{k}$. In the special case of normal dispersion, $n_{p}$ is independent of the angle of incidence . For the cases with dispersion close to normal $n_{p}$ vs. $\tilde{f}$ for up to the $6^{th}$ band has been calculated and the results can be seen in Fig. 1b. We observe in Fig. 1b that $|n_{p}|$ can have values less than one (i.e., superluminal phase velocity). However  $|n_{g}|$ has to be  always greater than one, because the group velocity is  the energy velocity in the PC, which is less than $c$. We have calculated the group velocity with the $\vec k \cdot \vec p$  perturbation method \cite{Busch} and checked that indeed $|n_{g}|>$1 . For the cases with normal dispersion  $\vec v_{g}$=$\frac{c}{n_{g}} \hat k$ with $n_{g}$=$\omega d|n_{p}|$/$d\omega+|n_{p}|$ \cite{proof}. From this relation and from the graph of $n_{p}$ versus frequency  (see Fig. 1b), it can be checked that indeed, $n_{g}>$0 when $n_{p}>$0 and $n_{g}<$0 when $n_{p}<$0. This means  that for the normal dispersion cases negative refraction occurs when the phase index is negative. We note this finding is different of that stated in Ref. 26. For the anisotropic dispersion cases $\vec v_{g}$ and  $\vec k$ are not along the same direction. In these cases we choose the sign of $n_{g}$ to manifests the sign of refraction at the interface and therefore  $n_{g}$ will have the sign of $(\vec v_{g} \cdot \vec k_{i})_{x}$ ({\it x} being the direction along the interface and $\vec k_{i}$ the incident wavevector). We would like to stress that $n_{g}$ cannot be used in a Snell-like formula to obtain the signal's propagation direction in the crystal. However, $n_{p}$ can be used in Snell's law to determine the signal's propagation angle in the normal dispersion case, under certain conditions \cite{fotein}.  
\par
There are two characteristic cases for which an EM wave refracts in the ``wrong way'' (negatively) when it hits the PC interface. One can predict the direction of the refracted signal from the equifrequency contours in the air and PC media. This methology has been used widely in 1D diffraction gratings\cite{confined}. The EFS for two frequencies, --$\tilde f=0.58$ and $\tilde f=0.535$ -- that are representative for these two cases, are seen in Fig. 2a and Fig. 2b. The $\vec k_{//}$ conservation condition given by the dashed vertical line in the figures determines the allowed refracted wavevector $\vec k_{f}$. In both cases there are two choices for $\vec k_{f}$ (to point either towards A or towards B). For every choosen $\vec k_{f}$, the corrresponding $\vec v_{g}$ will be perpendicular to the EFS at that point and point towards increasing values of frequency. The correct choice for $\vec k_{f}$ will be the one that gives a  $\vec v_{g}$ that points away from the source. After checking the EFS for a region around $\tilde f=0.58$ and $\tilde f=0.535$ we determine that the frequency increases inwards for the case of Fig. 2a (i.e. $\vec v_{g} \cdot \vec k <0$ and $n_{p}<$0 ) while it increases outwards for the case of Fig. 2b (i.e. $\vec v_{g} \cdot \vec k >0$ and $n_{p}>0$). Taking this into consideration the correct choice for $\vec k_{f}$ is to point towards A for the case of Fig. 2a and towards B for the case of Fig. 2b. For an infinite PC the group and the energy velocity are equal, so the group velocity vector (orange vector) in the figures represents the direction of propagation for the EM signal in the PC. We have confirmed the theoretical prediction of negative refraction with a FDTD simulation for both cases. In other words, we have determined both theoreticaly and numerically that  the group index $n_{g}$ for both the cases is negative.\cite{ngnote}. However this type of  FDTD simulation cannot decide what is the ``rightness'' of the system. 
\par  
Therefore we consider an incoming EM wave incident normally on a wedged structure, as was the case of the UCSD experiment \cite{shelbysc}. We should mention that couplings to higher order Bragg waves \cite{sakoda} when the wave hits the wedged interface are unavoidable in some cases. However, the sign of the angle of the $0^{th}$ order outgoing Bragg wave coincides with the sign of $\vec S \cdot \vec k$ inside the wedge, and determines the ``rightness''\cite{vesalago} (positive for the right and negative for the left handed case) and so the sign of $n_{p}$ of the PC. 
\par 
In particular, we show the results for this type of simulation for $\tilde f=0.58$ (case of Fig. 2a) and $\tilde f=0.535$ (case of Fig. 2b) in Figs 3b and 4b respectively. For each of these cases the symmetry directions of the first and the wedged interface are choosen appropriately \cite{fotein}.  Naturally, the $k_{//}$ wavevector will be conserved across both interfaces. In Fig. 3b the final outgoing beam is in the negative hemisphere (negative in respect to what is expected for a regular homogeneous material with positive index).  This means that the perpendicular to the first interface component of the wavevector has reversed sign when crossing the air PC interface. Since $\vec S$ is always pointing away from the source $\vec S \cdot \vec k $ will be negative in the PC for the case of Fig. 3b. We note that the second outgoing beam in Fig. 3b is due to consequent multi-reflection on the upper corner (one can determine that by looking at the field inside the wedge). To extract the sign of $\vec S \cdot \vec k$ in Fig. 4b, because of the presence of the $2^{nd}$ beam (higher order Bragg wave \cite{proof2}) we need the additional knowledge of the theoretically expected magnitude of the angle for the $0^{th}$ order Bragg wave. We use the EFS contours for the infinite system to determine the magnitude of the refracted wavevector and from that the magnitude of the $0^{th}$ order Bragg wave. (We obtained a value of $57^0$). Although the actual magnitude of the refracted wavevector  may differ from the predicted one, no dramatic difference is expected. We then observe   
in the FDTD simulation of Fig. 4b that the beam with deflection angle of magnitude close to $57^0$ lies in the positive hemisphere. So, in the case of Fig. 4b $\vec S \cdot \vec k >0$.  
\par
The ray schematics in Figs 3a and 4a -- that accompany the simulations of Figs. 3b and Fig. 4b-- show the theoretically predicted refracted and final outgoing beams by using the properties of the infinite system. The beams are determined in a similar manner as in the ray schematics of Fig. 2. In the following we compare the theoretical expected angles (magnitude and sign) with the ones measured in the FDTD simulation.  The expected outgoing angle for the case of Fig. 3b is $-36^0$, while the FDTD one is $\sim -30^0$. Likewise for  the case of Fig. 4 the formula in Ref. 28 predicts an outgoing wave at  $57^0$ and a $1^{st}$ order Bragg wave at $-14^0$. These values are close to those of  $\sim 60^0$ and $\sim -10^0$ measured in the FDTD simulation. 
\par  
By looking on other bands we can find cases with $n_{p}>0$ and $n_{g}>0$, as well as $n_{p}<0$ and $n_{g}>0$ , so all combinations of signs for the group and phase index are possible. By performing FDTD simulations (of both types (slab and wedge)) we have shown that in all cases, that the ``rightness'' of the PC -- i.e. the sign of $\vec S \cdot \vec k$ \cite{note}--, coincides with the the theoretical prediction for the sign of $n_{p}$ , --as defined earlier in the paper for the infinite PC--, and is irrelevant to the sign of $n_{g}$. For the case of Fig. 2a, the PC is left-handed and the perpendicular component of the wavevector reverses sign in the air PC interface. In addition it has close to normal dispersion and a negative effective index. So one could say that the negative refraction in this case is similar to the one occurring in a regular LH material that has both $\epsilon$ and $\mu$ negative.\cite{smithpbg,smithprl,shelbyapl,shelbysc,mehmet,vesalago,pendry,pendry00,markos,smithprb}  
\par    
In conclusion, we have systematically examined the conditions of obtaining left-handed behavior in photonic crystals. We have demonstrated that the existence of negative refraction is neither a prerequisite nor guarantees  a negative effective index and thus LH behavior. Although we have focused on the presentation of 2D photonic crystals, the results are general enough and are expected to be correct for 3D PCs. Special care must be taken for the couplings of the TE modes and TM modes in the 3D systems.  We have suggested experiments to check our theoretical predictions both in the microwave regime as well as in the optical regime. The present work will help in the understanding and design of the optical devices.  
 \par   
\acknowledgments   
We thank K. Busch for useful discussions regarding the group velocity calculation with the $\vec k \cdot \vec p$ perturbation method. We would also like to thank Mario Agio for useful discussions. Ames Laboratory is operated by the U.S Department of Energy by Iowa State University under Contract No. W-7405-Eng-82. This work was partially supported by DARPA.

\begin{figure}[b] 

\includegraphics[angle=270,width=3.50cm]{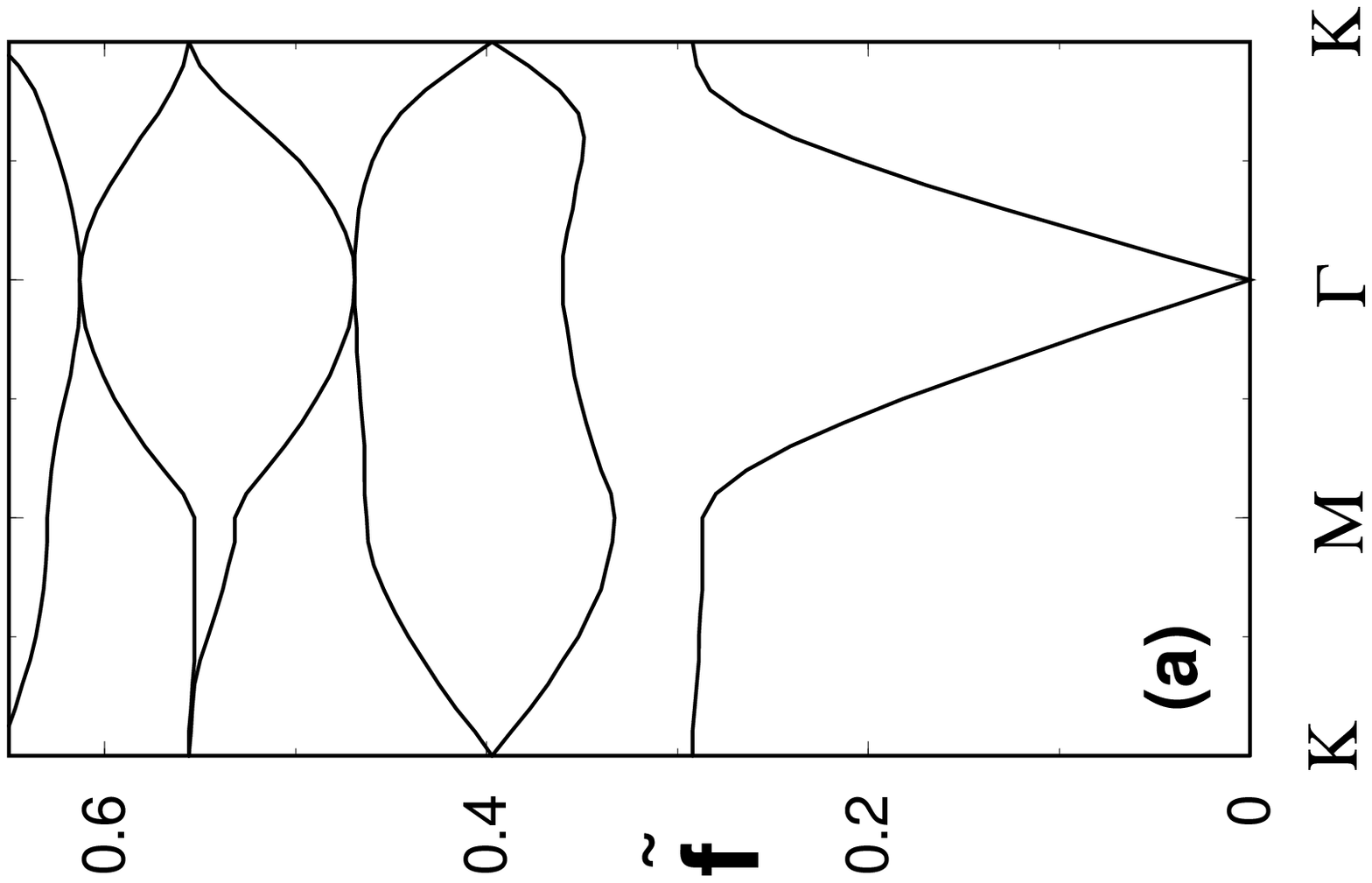}
\includegraphics[angle=270,width=3.05cm]{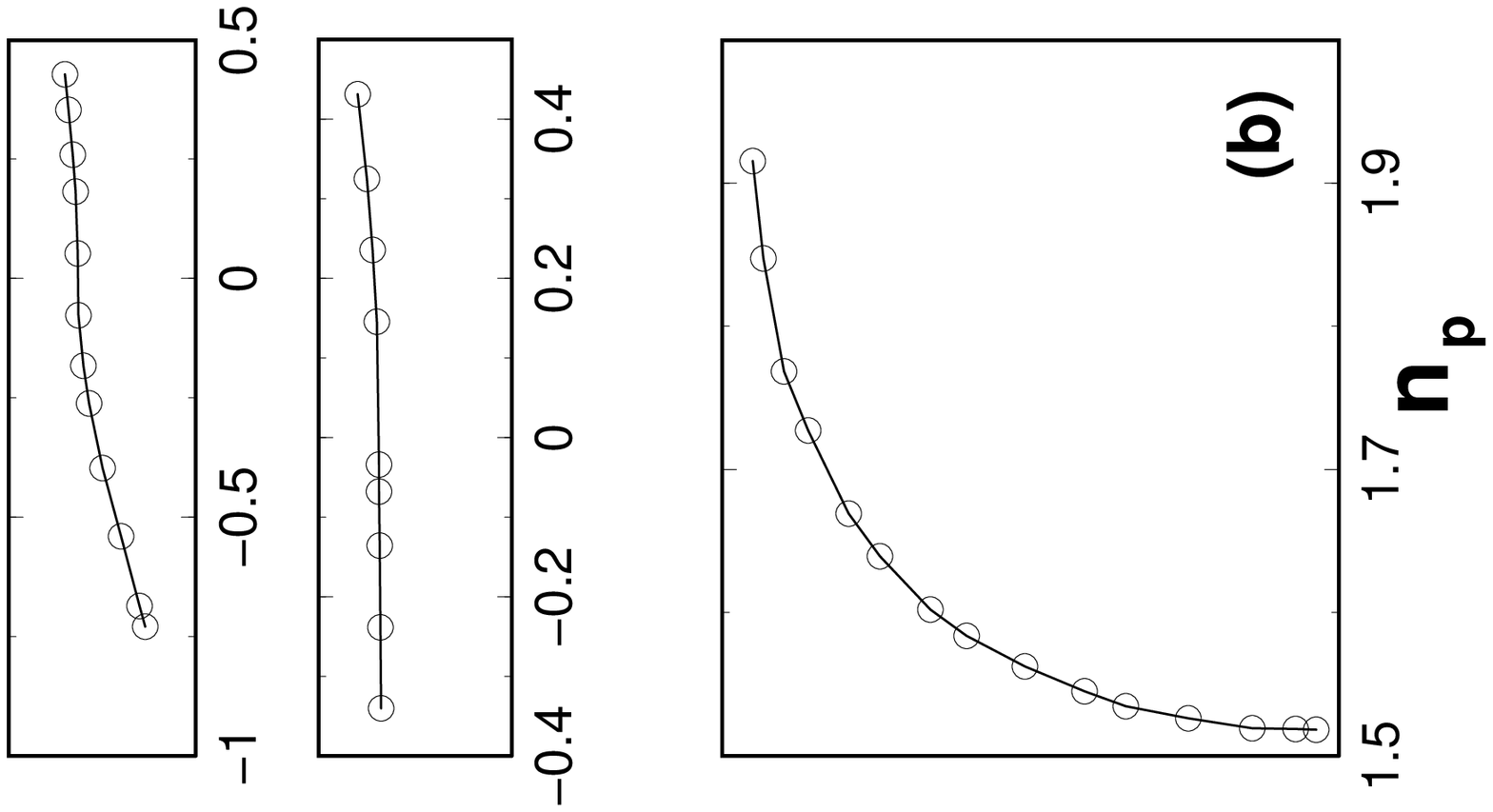} 

\caption{\label{Figure 1} a) The band structure for the H (TE) polarization for the PC structure under study. b) The effective phase index $n_{p}$ plotted for the frequency regions where the dispersion is close to normal.}  
\end{figure}

\begin{figure}[!h] 
\includegraphics[angle=0,width=6.7cm,height=3.75 cm]{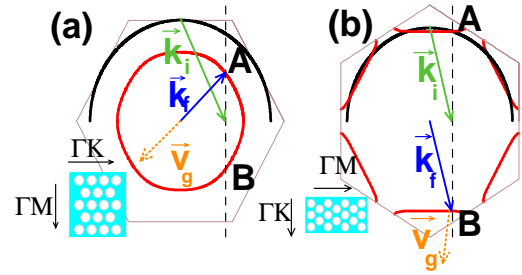}
\caption{\label{Figure 2} Schematics of refraction at the PC interface for the four different cases. The initial ($\vec {k}_{i}$) and refracted ($\vec {k}_{f}$) wave vectors as well as the group velocity ($\vec {v}_{g}$) of the refracted wave are drawn. The dashed line represents the conservation of the parallel component of the wave vector. } 
\end{figure}
 
\begin{figure}[h]   
\begin{flushleft}
\includegraphics[angle=0,width=3.5cm]{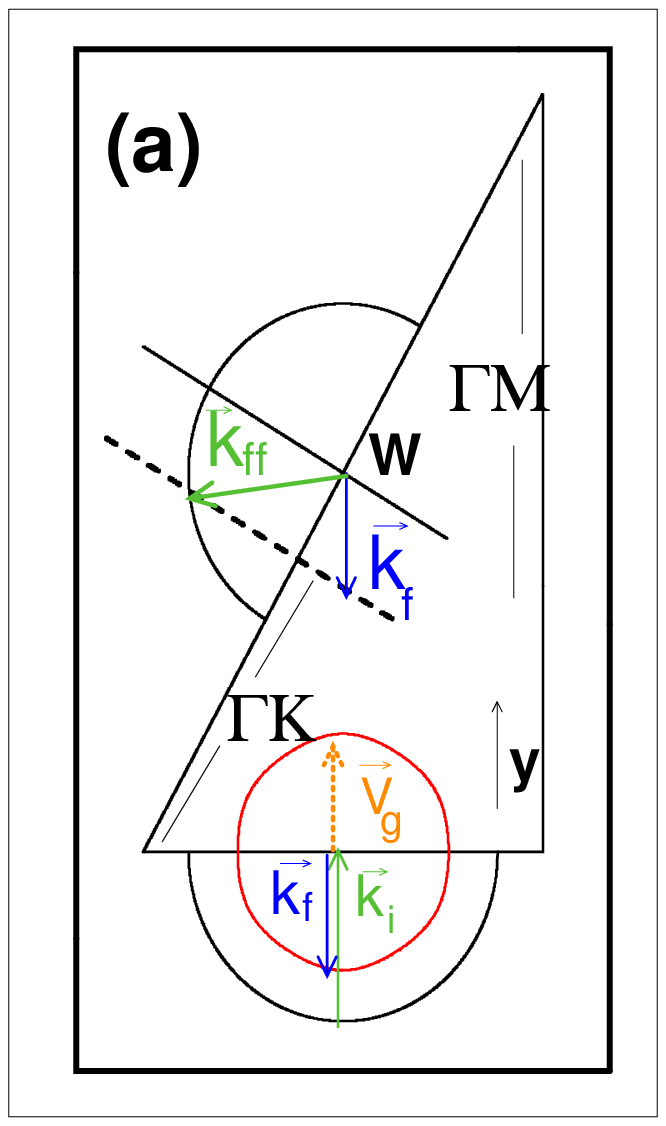}
\includegraphics[angle=0,width=3.7cm]{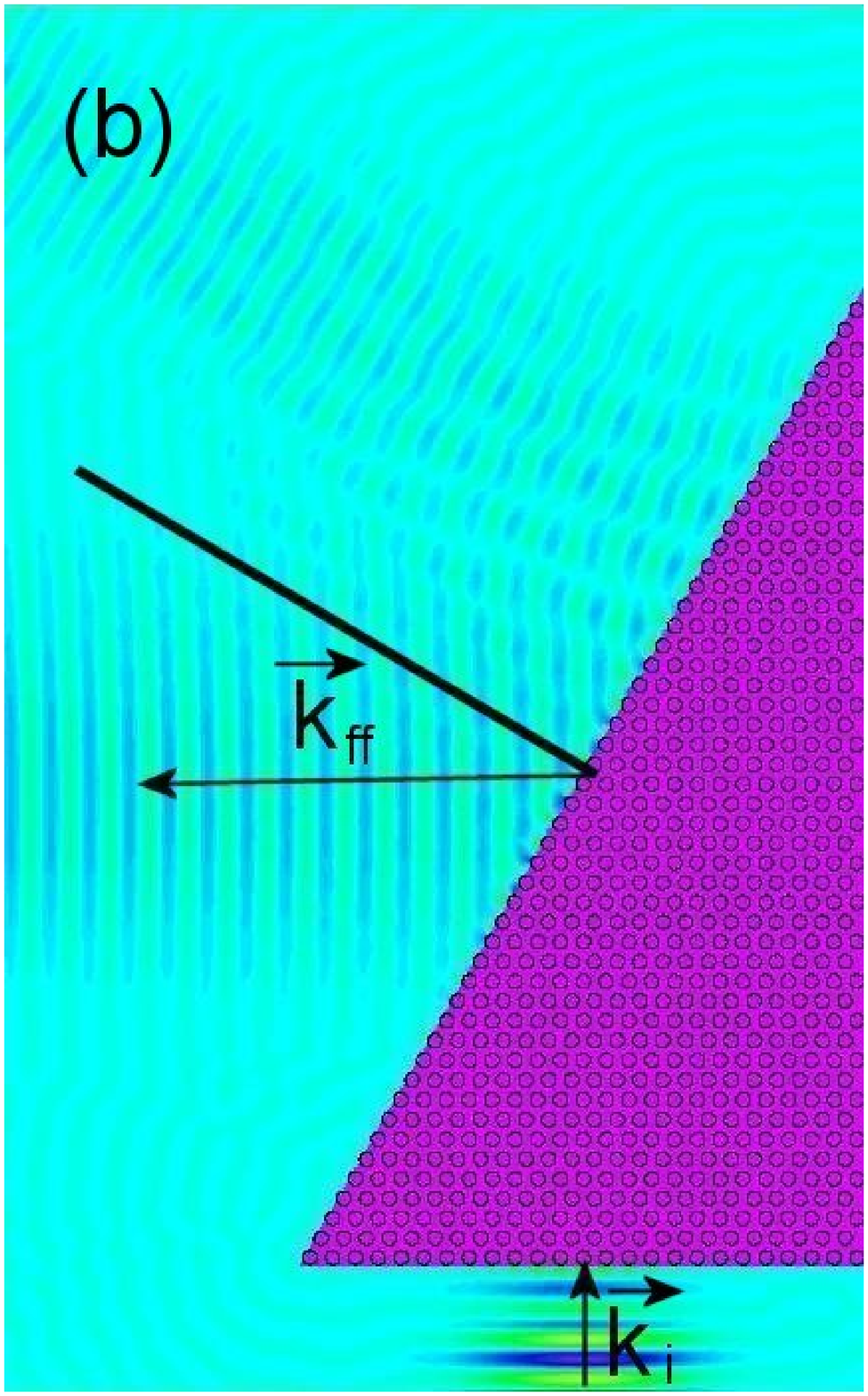}
\end{flushleft}
\caption{\label{Figure 3} Refraction of monochromatic EM wave with $\tilde f=0.58$ incident normally on a PC crystal wedge. In (a) the expected ray-paths are shown. The wave vectors of the incident ($\vec k_{i}$) as well as the refracted ($\vec k_{f}$) and the outgoing ($\vec k_{ff}$) EM waves are shown. In (b) the actual FDTD simulation with the incident and outgoing beam is shown. The bold solid line is the normal to the wedged interface.}
\end{figure}
 
\begin{figure}[h]  

\includegraphics[angle=0,width=3.5cm]{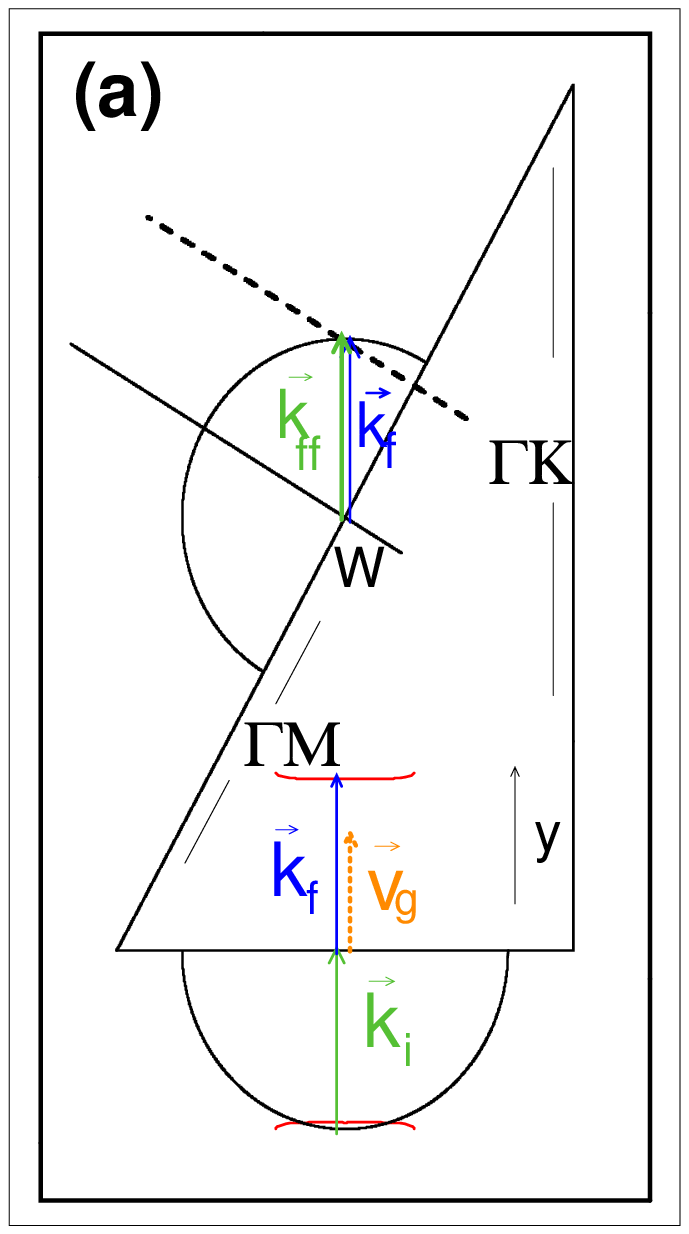}
\includegraphics[angle=0,width=3.3cm]{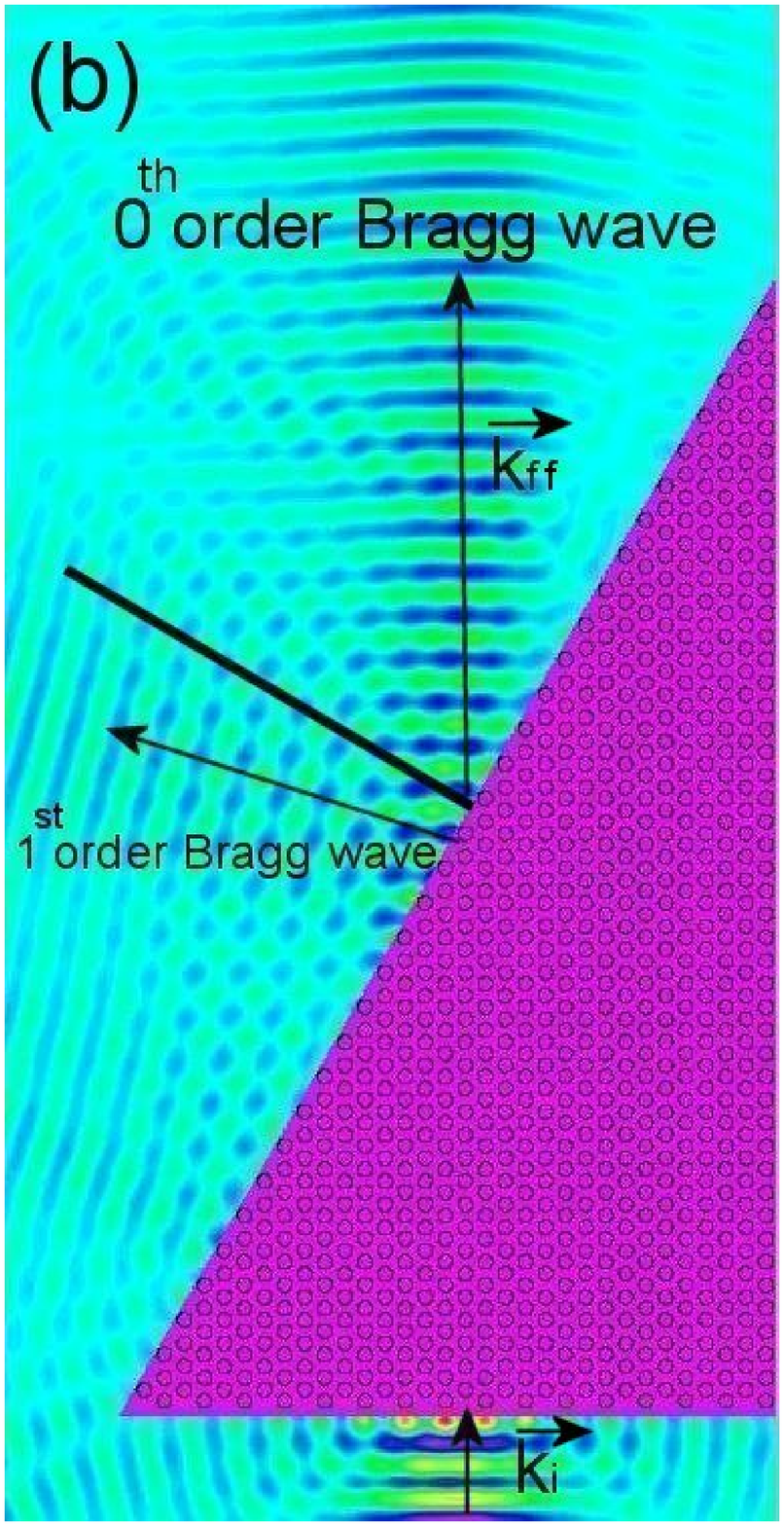}

\caption{\label{Figure 4} The same as figure 3 but for $\tilde f=0.535$}
\end{figure}

\clearpage
 

\begin{thebibliography}{710} 
\bibitem{smithpbg}  
For a recent review see the articles of D. R. Smith and J. Pendry in  {\it Photonic crystals and Light Localization in the 21st Century} ed. by C. M. Soukoulis, NATO Science series C vol. 563, Kluwer-Dordrecht (2001), pp 329 and 351.
\bibitem{smithprl} 
D. R. Smith, W. J. Padilla, D. C. Vier, S. C. Nemat-Nasser, and S. Schultz, Phys. Rev. Lett. {\bf 84}, 4184 (2000).
\bibitem{shelbyapl} 
R. A. Shelby, D. R. Smith, S. C. Nemat-Nasser and S. Schultz, Appl. Phys. Lett. {\bf 78}, 489 (2001)  
\bibitem{shelbysc} 
R. A. Shelby, D. R. Smith and S. Schultz, Science {\bf 77}, 292 (2001).
\bibitem{mehmet}
M. Bayindir et. al., Appl. Phys. Lett. {\bf 81}, 120 (2002)
\bibitem{vesalago}   
V. G. Vesalago, Usp. Fiz. Nank {\bf 92}, 517 (1968) (Sov. Phys. Usp. {\bf 10}, 509 (1968))

\bibitem{pendry}
J. B. Pendry, A. J. Holden, W. J. Stewart, and I. Youngs, Phys. Rev. Lett. {\bf 76}, 4773 (1996); J B Pendry, A. J.  Holden, D. J. Robbins and W. J. Stewart, J. Physics: Condens. Matt. {\bf 10}, 4785 (1998); J. B. Pendry, A. J. Holden, D.J. Robbins, W.J and Stewart, IEEE Trans. Microwave Theory Tech. {\bf 47}, 2075 (1999).

\bibitem{pendry00}
J. B. Pendry, Phys. Rev. Lett. {\bf 85}, 3966 (2000).

\bibitem{markos}
P. Marko\v{s} and C. M. Soukoulis, Phys. Rev B {\bf 65}, 033401 (2001); P. Marko\v{s} and C. M. Soukoulis, Phys. Rev. E {\bf 65}, 036622 (2002).

\bibitem{smithprb}
D. R. Smith, S. Schultz, P. Marko\v{s} and C. M. Soukoulis, Phys. Rev. B {\bf 65}, 195104 (2002).

\bibitem{Notomi}
M. Notomi, Phys. Rev. B {\bf 62},10696 (2000).

\bibitem{kosaka}
H. Kosaka {\it et al.}, Phys. Rev. B {\bf  58},R10096 (1998).
 
\bibitem{Russel} 
P. St. J. Russel, Phys. Rev. A {\bf 33}, 3232 (1986); P. St. Russell and T. A. Birks in {\it Photonic Band Gap Materials}  ed. by C. M. Soukoulis , NATO ASI series E, vol. 315, Kluwer-Dordrecht (1996), pp71.

\bibitem{Luo}   C. Luo, S. G. Johnson, J. D. Joannopoulos,
J. B. Pendry, Phys. Rev. B. {\bf 65}, 201104(R) (2002).

\bibitem{Gralak}
B. Gralak, S. Enoch and G. Tayed, J. of Opt. Soc. Amer. A {\bf  17}, 1012 (2000).
\bibitem{note}
The PC structure is periodic in space. Therefore whenever we refer to $\vec S$ in the PC the spatial average within the unit cell of the time-averaged Poynting vector is considered.
\bibitem{tavlove}
A. Taflove, {\it Computational Electrodynamics - The Finite Difference Time-Domain Method } (Artech House, 1995); K. S. Yee, IEEE trans. on Antennas and Propagation ap-14, 302 (1966).
\bibitem{Berenger}
J. P. Berenger, J. of Comp. Phys. {\bf 114},185 (1994); J. P. Berenger, IEEE Trans. on Antennas and Propagation,{\bf 44}, 110 (1996);
\bibitem{chan}
K.M. Ho, C. T. Chan and C. M. Soukoulis, Phys. Rev. Let.  {\bf 65}, 3152 (1990).
\bibitem{villen} 
P. R. Villeneuve and. M. Piche, Prog. of Quant. Electr. {\bf 18}, 153 (1994).
\bibitem{fotein}
S. Foteinopoulou and C. M. Soukoulis, unpublished.
\bibitem{sakoda}
K. Sakoda {\it Optical properties of photonic crystals}, Springer, 2001.
\bibitem{Busch}
D. Hermann, M. Frank, K. Busch and P. W\"olfle, Optics Express {\bf 8}, 167 (2001)
\bibitem{proof}
$|\vec k|=\sqrt{k_{x}^2+k_{y}^2}=R(\omega)$ (where $R$ the radius of the circular EFS). Since $|\vec v_{p}|=\omega/|\vec k|=c/|n_{p}|$, $\omega=cR(\omega)/|n_{p}|$. So $\vec v_{g}=\vec \nabla_{k} \omega=\hat k/(dR/d\omega)=c \hat k/(\omega d|n_{p}|/d\omega+|n_{p}|)$
\bibitem{confined}  
P. St. Russel, T. A. Birks and F. Dominic Lloyd-Lucas in ``{\it Confined Electrons and Photons, New Physics and Applications}'', ed. E. Burstein and C. Weisbuch, NATO ASI series Vol. 340, Plenum Press, New York, 1995, p. 585. 
\bibitem{walser}
R. M. Valanju, R. M. Walser and A. P. Valanju, Phys. Rev. Lett. {\bf 88}, 187401 (2002). 
\bibitem{ngnote} 
$n_{g}$ for the case of Fig. 2b is a function of the angle of incidence $\theta$. The sign quoted in the text for $n_{g}$ for this case is applicable for a certain angle span around the normal direction.
\bibitem{proof2} 
The outgoing  angle  $\theta_{out,m}$ is: $\tan(\theta_{\rm out,m})$=$\frac{k_{xp,m}}{\sqrt{(\omega^2/c^2)-k_{xp,m}^2}} $
with $ k_{xp,m}=k_{fy} \cos(\theta_{w})+\frac{2 m \pi}{b}$
where $m$ is the order of the Bragg wave, $\theta_{w}=30^0$ and $b=a$ for Fig.3 and $\sqrt{3} a$ for Fig.4. $k_{fy}$ is the projection of the wavevector in the wedged structure along the y- direction (see Figs. 3a and 4a).\\

\end{thebibliography}
\end{document}